\documentclass[showpacs,preprintnumbers,amsmath,amssymb,nofootinbib]{revtex4}

\usepackage{graphicx}
\usepackage{dcolumn}
\usepackage{bm}
\usepackage{amssymb}
\usepackage{amsfonts}
\usepackage{amsmath}
\usepackage{latexsym}
\usepackage{dsfont}
\usepackage{multirow}
\usepackage{color}
\usepackage[mathscr]{eucal}

\definecolor{nv}{rgb}{0.1,0.1,0.6}
\definecolor{pr}{rgb}{0.2,0.1,0.5}
\definecolor{mg}{rgb}{0.4,0.0,0.4}

\newcommand{\beq}{\begin{equation}}
\newcommand{\eeq}{\end{equation}}
\newcommand{\beqy}{\begin{eqnarray}}
\newcommand{\eeqy}{\end{eqnarray}}
\newcommand{\beqyn}{\begin{eqnarray*}}
\newcommand{\eeqyn}{\end{eqnarray*}}

\newcommand{\bs}{\begin{slide}}
\newcommand{\es}{\end{slide}}
\newcommand{\bc}{\begin{center}}
\newcommand{\ec}{\end{center}}
\newcommand{\bmin}{\begin{minipage}}
\newcommand{\emin}{\end{minipage}}

\newcommand{\bi}{\begin{itemize}}
\newcommand{\ei}{\end{itemize}}

\newcommand{\bea}{\begin{eqnarray}}
\newcommand{\eea}{\end{eqnarray}}
\newcommand{\be}{\begin{equation}}
\newcommand{\ee}{\end{equation}}

\newcommand{\ud}{\mathrm{d}}

\newcommand{\uTr}{\mathrm{Tr}}

\newcommand{\uslash}{/\!\!\!}

\newcommand{\uD}{\mathcal{D}}

\newcommand{\barpsi}{\overline{\psi}}

\newlength\savedwidth

\newcommand\whline{\noalign{\global\savedwidth\arrayrulewidth
\global\arrayrulewidth 1pt}%
\hline
\noalign{\global\arrayrulewidth\savedwidth}}

\newcommand{\pure}{\text{pure}}
\newcommand{\phys}{\text{phys}}

\newcommand{\LRpartial}{\overset{\leftrightarrow}{\partial}\!\!\!\!\phantom{\partial}}
\newcommand{\LRD}{\overset{\leftrightarrow}{D}\!\!\!\!\!\phantom{D}}

\begin{document}


\title{The light-front gauge-invariant energy-momentum tensor}

\author{C\'edric Lorc\'e}
\email{clorce@slac.stanford.edu;C.Lorce@ulg.ac.be}
\affiliation{SLAC National Accelerator Laboratory, Stanford 
  University, Menlo Park, CA 94025 USA}
\affiliation{IFPA,  AGO Department, Universit\'e de Li\` ege, Sart-Tilman, 4000 Li\`ege, Belgium}

\date{\today}

\begin{abstract}
We provide for the first time a complete parametrization for the matrix elements of the generic asymmetric, non-local and gauge-invariant canonical energy-momentum tensor, generalizing therefore former works on the symmetric, local and gauge-invariant kinetic energy-momentum tensor also known as the Belinfante-Rosenfeld energy-momentum tensor. We discuss in detail the various constraints imposed by non-locality, linear and angular momentum conservation. We also derive the relations with two-parton generalized and transverse-momentum dependent distributions, clarifying what can be learned from the latter. In particular, we show explicitly that two-parton transverse-momentum dependent distributions cannot provide any model-independent information about the parton orbital angular momentum. On the way, we recover the Burkardt sum rule and obtain similar new sum rules for higher-twist distributions.
\end{abstract}

\pacs{11.15.-q,12.38.Aw,13.88.+e,13.60.Hb,14.20.Dh}
\maketitle

\section{Introduction}

Following the canonical procedure based on Noether's theorem, one ends up with a canonical energy-momentum tensor which is usually neither symmetric nor gauge invariant. Because of these pathologies, one often abandons the canonical energy-momentum tensor in favor of the Belinfante-Rosenfeld improved energy-momentum tensor~\cite{Belinfante:1939,Belinfante:1940,Rosenfeld:1940} which is both symmetric and gauge invariant. The Belinfante-Rosenfeld tensor differs from the canonical tensor by a so-called superpotential term which modifies the definition the momentum density but leaves both the \emph{total} linear and angular momenta unchanged. It has the peculiar feature that it denies the mere existence of spin density. Indeed, from the conservation of total angular momentum $\partial_\mu J^{\mu\nu\rho}=0$ where $J^{\mu\nu\rho}=r^\nu T^{\mu\rho}-r^\rho T^{\mu\nu}+S^{\mu\nu\rho}$ with $T^{\mu\nu}$ the conserved total energy-momentum tensor and $S^{\mu\nu\rho}$ the spin density tensor, one deduces that the antisymmetric part of the energy-momentum tensor is intimately related to the quark spin density $T^{\nu\rho}-T^{\rho\nu}=-\partial_\mu S^{\mu\nu\rho}$. So, in the Belinfante-Rosenfeld approach, what is usually refered to as ``spin'' is simply described as a flow of momentum. There is therefore no clear distinction between spin and orbital angular momentum (OAM) in this approach, just like there is no clear distinction between energy flow $T^{i0}$ and momentum density $T^{0i}$.

On the other hand, spin is an intrinsic property of a particle defined as one of the two Casimir invariants of the Poincar\'e group (the other Casimir invariant being the mass). Contrary to OAM, one cannot change the spin of a particle by changing the Lorentz frame. Spin and OAM are distinguishable, and so dealing with a symmetric energy-momentum tensor is not very natural in Particle Physics. Where does this symmetry requirement come from? It is mainly motivated by General Relativity where gravity couples to a symmetric energy-momentum tensor. It is however important to notice that General Relativity is a classical theory while spin is fundamentally a quantum concept. Moreover, the symmetry of the energy-momentum tensor in General Relativity follows from the postulated absence of space-time torsion. More general theories relax the no-torsion assumption and do not require the energy-momentum tensor to be symmetric. The gravitational effects of the antisymmetric part of the energy-momentum tensor are however extremely small and are expected to show up only under extreme conditions, see \emph{e.g.}~\cite{Hehl:1976kj,Hehl:1994ue,Obukhov:2014fta} and references therein. Finally, we note that the classical argument in favor of a symmetric energy-momentum tensor based on dimensional analysis and presented \emph{e.g.} in section 5.7 of~\cite{Misner:1974qy}, is valid only for the orbital form of angular momentum.
\newline

The early papers about the proton spin decomposition~\cite{Jaffe:1989jz,Ji:1996ek,Shore:1999be} start with the Belinfante-Rosenfeld tensor, but then add appropriate superpotential terms to decompose the quark angular momentum into spin and orbital contributions. According to textbooks~\cite{Cohen,Simmons}, no such decomposition is possible for the gauge field angular momentum. Though, photon spin and OAM are routinely measured in Quantum ElectroDynamics, see \emph{e.g.}~\cite{Bliokh:2014ara} and references therein. In Quantum ChromoDynamics (QCD), a quantity called $\Delta G$ which can be interpreted in the light-front gauge $A^0+A^3=0$ as the gluon spin~\cite{Jaffe:1989jz} has been measured in polarized deep inelastic and proton-proton scatterings, see~\cite{deFlorian:2014yva} for a recent analysis. In order to account for these experimental facts, Chen \emph{et al.} claimed in 2008 that the textbooks were wrong, and proposed a formal gauge-invariant decomposition of the photon and gluon angular momentum~\cite{Chen:2008ag}. This triggered a lot of criticism and an outpouring of theoretical papers, summarized in the recent reviews~\cite{Leader:2013jra,Wakamatsu:2014toa}. The apparent contradiction with the textbook claim was solved by realizing that the Chen \emph{et al.} construction is intrinsically non-local~\cite{Hatta:2011zs,Lorce:2012rr,Lorce:2012ce}, whereas textbooks implicitly refered to local quantities only. It has actually been known for quite some time that gauge invariance can be restored by allowing the quantities to be non-local~\cite{Dirac:1955uv,Mandelstam:1962mi}. Although there are in principle infinitely many gauge-invariant non-local quantities reducing formally to the same gauge non-invariant local expression in the appropriate gauge, the experimental conditions ultimately determine which ones are accessible~\cite{Lorce:2013bja}.

Parton distributions are typical examples of measurable non-local quantities. Gauge invariance is ensured by a Wilson line whose path is determine by the factorization theorems~\cite{Collins:2011zzd}. Ji has shown that the kinetic OAM, which is local and gauge invariant, can be expressed in terms of Generalized Parton Distributions (GPDs) that are accessible in some exclusive experiments like \emph{e.g.} Deeply Virtual Compton Scattering~\cite{Ji:1996ek}. Since the local expression for the canonical OAM is gauge non-invariant~\cite{Jaffe:1989jz}, it was thought for a long time that it cannot be measured and should therefore be considered as unphysical. The situation has changed once it has been realized that the non-local expression for the canonical OAM, which is gauge invariant, can be expressed in terms of $k_T$-dependent GPDs, also known as Generalized Transverse-Momentum dependent Distributions (GTMDs)~\cite{Lorce:2011kd,Lorce:2011ni,Hatta:2011ku}. GTMDs are extremely interesting as they provide the maximal information about the phase-space or Wigner distribution of quarks and gluons. Unfortunately, apart possibly in the low-$x$ regime, it is not known so far how to access these GTMDs experimentally~\cite{Meissner:2009ww}. The situation is however not hopeless since GTMDs can be accessed indirectly using realistic models, see \emph{e.g.}~\cite{Lorce:2011kd,Kanazawa:2014nha,Mukherjee:2014nya,Mukherjee:2015aja,Liu:2014vwa,Miller:2014vla,Liu:2015eqa}. Another possibility is to compute the GTMDs on the lattice. The traditional approach is to compute moments of the parton distributions using a tower of gauge-invariant local operators. Unfortunately, this approach does not allow one to compute $\Delta G$ because the latter does not correspond to any gauge-invariant local operator. However, new strategies have recently been proposed allowing in principle the computation of matrix element of non-local operators on the lattice, and show already encouraging results~\cite{Musch:2010ka,Musch:2011er,Ji:2013dva,Ji:2014lra,Lin:2014zya,Ma:2014jla}.
\newline

\begin{table}[t!]
\begin{center}
\caption{\footnotesize{Properties of the various forms of the energy-momentum tensor in a gauge theory.}}\label{EMT}
\begin{tabular}{@{\quad}c@{\quad}c@{\quad}|@{\quad}c@{\quad}c@{\quad}c@{\quad}}\whline
Family&Energy-momentum tensor&Gauge invariant&Local&Symmetric\\
\hline
&Belinfante-Rosenfeld~\cite{Belinfante:1939,Belinfante:1940,Rosenfeld:1940}&$\checkmark$&$\checkmark$&$\checkmark$\\
Kinetic&Ji~\cite{Ji:1996ek}&$\checkmark$&$\checkmark$&$-$\\
&Wakamatsu~\cite{Wakamatsu:2010qj}&$\checkmark$&$-$&$-$\\
\hline
\multirow{2}{*}{Canonical}&Jaffe-Manohar~\cite{Jaffe:1989jz}&$-$&$\checkmark$&$-$\\
&Chen \emph{et al.}~\cite{Chen:2008ag}&$\checkmark$&$-$&$-$\\
\whline
\end{tabular}
\end{center}
\end{table}

Many different forms have been proposed for the energy-momentum tensor in a gauge theory. Their properties are summarized in Table~\ref{EMT}. All the forms can be sorted into two families~\cite{Wakamatsu:2010qj,Wakamatsu:2010cb,Lorce:2012rr} : kinetic (or mechanical) and canonical. They all give the same total linear momentum, but attribute different momentum densities to the various constituents. To the best of our knowledge, only the matrix elements of the local energy-momentum tensors have been discussed in the literature so far. The first complete parametrization of the matrix elements of the symmetric gauge-invariant local operator (\emph{i.e.} Belinfante-Rosenfeld tensor) has been given in~\cite{Ji:1996ek} and further discussed in~\cite{Polyakov:2002yz}. The matrix elements of the asymmetric local gauge-invariant operator (\emph{i.e.} Ji tensor) has first been discussed in~\cite{Shore:1999be}, but the correct parametrization in the off-forward case was given in~\cite{Bakker:2004ib}. There has also been a simple attempt to parametrize in a similar way the matrix elements of the asymmetric local gauge non-invariant operator (\emph{i.e.} Jaffe-Manohar tensor), but this led to the absurd conclusion that canonical and kinetic matrix elements are the same~\cite{Leader:2012vp}. We will argue in the present paper that the failure of this attempt can be understood as due to the absence of an important piece of information in the parametrization of~\cite{Leader:2012vp}. We will also show explicitly that two-parton Transverse-Momentum Distributions (TMDs), though sensitive to OAM, cannot provide any quantitative model-independent information about the OAM.
\newline

The paper is organized as follows. In section~\ref{sec2}, we decompose the QCD energy-momentum and generalized angular momentum tensors into quark and gluon contributions, and compare the various forms found in the literature. In section~\ref{sec3}, we provide for the first time the parametrization of the generic non-local light-front gauge-invariant energy-momentum tensor and discuss various constraints in section~\ref{sec4}. In section~\ref{sec5}, we derive the relations between the scalar functions appearing in this parametrization and derive the relations with the two-parton generalized and transverse-momentum dependent distributions, obtaining on the way new sum rules. Finally, we gather our conclusions in section~\ref{sec6}. Some details about the parametrization are given in Appendix~\ref{App}

\section{The gauge-invariant linear and angular momentum tensors}\label{sec2}

In order to deal most conveniently with the various gauge-invariant decompositions proposed in the literature, we consider the following five gauge-invariant energy-momentum tensors
\begin{equation}
\begin{aligned}
T^{\mu\nu}_1(r)&=\barpsi(r)\gamma^\mu \tfrac{i}{2}\LRD^\nu\psi(r),\\
T^{\mu\nu}_2(r)&=-2\uTr\!\left[G^{\mu\alpha}(r)G^\nu_{\phantom{\nu}\alpha}(r)\right]+g^{\mu\nu}\,\tfrac{1}{2}\uTr\!\left[G^{\alpha\beta}(r)G_{\alpha\beta}(r)\right]\!,\\
T^{\mu\nu}_3(r)&=-\barpsi(r)\gamma^\mu gA^\nu_\phys(r)\psi(r),\\
T^{\mu\nu}_4(r)&=\tfrac{1}{4}\,\epsilon^{\mu\nu\alpha\beta}\partial_\alpha\!\left[\barpsi(r)\gamma_\beta\gamma_5\psi(r)\right]\!,\\
T^{\mu\nu}_5(r)&=-2\partial_\alpha\uTr\!\left[G^{\mu\alpha}(r)A^\nu_\phys(r)\right]\!,
\end{aligned}
\end{equation}
where $\epsilon_{0123}=+1$ and $\tfrac{i}{2}\LRD^\mu=\tfrac{i}{2}\LRpartial^\mu+gA^\mu$ is the hermitian covariant derivative with $\LRpartial^\mu=\overset{\rightarrow}{\partial}\!\!\!\!\phantom{\partial}^\mu-\overset{\leftarrow}{\partial}\!\!\!\!\phantom{\partial}^\mu$. Similarly, we consider the following seven gauge-invariant generalized angular momentum tensors
\begin{equation}\label{AMtensors}
\begin{aligned}
L^{\mu\nu\rho}_a(r)&=r^\nu T^{\mu\rho}_a(r)-r^\rho T^{\mu\nu}_a(r),\qquad a=1,\cdots,5,\\
S^{\mu\nu\rho}_1(r)&=\tfrac{1}{2}\,\epsilon^{\mu\nu\rho\sigma}\,\barpsi(r)\gamma_\sigma\gamma_5\psi(r),\\
S^{\mu\nu\rho}_2(r)&=-2\uTr\!\left[G^{\mu[\nu}(r)A^{\rho]}_\phys(r)\right]\!,
\end{aligned}
\end{equation}
where $x^{[\mu}y^{\nu]}=x^\mu y^\nu-x^\nu y^\mu$. The standard expressions for the Belinfante-Rosenfeld, Ji, Wakamatsu and Chen \emph{et al.} decompositions\footnote{We used the original covariant form of Ref.~\cite{Wakamatsu:2010cb} and not the one in Ref.~\cite{Leader:2013jra} which differs only by how one separates the pure-boost terms into quark and gluon contributions.} are then obtained by combining these contributions according to Tables~\ref{EMTdec} and~\ref{AMTdec}, and using the following identities based on the QCD equations of motion
\begin{equation}\label{QCDEOM}
\begin{aligned}
\barpsi(r)\gamma^{[\mu} i\LRD^{\nu]}\psi(r)&=-\epsilon^{\mu\nu\alpha\beta}\partial_\alpha\!\left[\barpsi(r)\gamma_\beta\gamma_5\psi(r)\right]\!,\\
2\!\left[\uD_\alpha G^{\alpha\beta}(r)\right]^c_{\phantom{c}c'}&=-g\,\barpsi_{c'}(r)\gamma^\beta\psi^c(r),
\end{aligned}
\end{equation}
where $c,c'$ are color indices in the fundamental representation and $\mathcal D_\mu=\partial_\mu-ig[A_\mu,\quad]$ is the adjoint covariant derivative. In particular, because of the first identity in Eq.~\eqref{QCDEOM}, we can write $T^{\mu\nu}_4(r)=-\tfrac{1}{2}\,T^{[\mu\nu]}_1(r)$ and therefore discard the tensor $T^{\mu\nu}_4(r)$ in the following discussions. Note that the tensors $T^{[\mu\nu]}_1(r)$, $T^{\mu\nu}_5(r)$, $L^{\mu\nu\rho}_4(r)-S^{\mu\nu\rho}_1(r)$ and $L^{\mu\nu\rho}_5(r)+S^{\mu\nu\rho}_2(r)$ have the form of a superpotential $\partial_\alpha f^{[\alpha\mu]\cdots}(r)$~\cite{Jaffe:1989jz}. Assuming as usual that surface terms vanish, this means that we have
\begin{equation}\label{antisymm}
\begin{aligned}
\partial_\mu T^{[\mu\nu]}_1(r)&=0,&\int\ud^3r\,T^{[n\nu]}_1(r)&=0,\\
\partial_\mu T^{\mu\nu}_5(r)&=0,&\int\ud^3r\,T^{n\nu}_5(r)&=0,\\
T^{[\nu\rho]}_1(r)&=-\partial_\mu S^{\mu\nu\rho}_1(r),&\qquad\qquad\int\ud^3r\,L^{n\nu\rho}_4(r)&=\int\ud^3r\,S^{n\nu\rho}_1(r),\\
T^{[\nu\rho]}_5(r)&=-\partial_\mu S^{\mu\nu\rho}_2(r),&\int\ud^3r\,L^{n\nu\rho}_5(r)&=-\int\ud^3r\,S^{n\nu\rho}_2(r),
\end{aligned}
\end{equation}
where $n$ is a timelike or lightlike four-vector and $\ud^3r=\epsilon_{\alpha\beta\gamma\delta}\,n^\alpha\,\ud r^\beta\wedge\ud r^\gamma\wedge\ud r^\delta$ is the volume element. This ensures that the quark and gluon linear and angular momenta are the same in the three kinetic decompositions
\begin{equation}
\begin{aligned}
\int\ud^3r\,T^{n\nu}_{\text{Bel},a}(r)&=\int\ud^3r\,T^{n\nu}_{\text{Ji},a}(r)=\int\ud^3r\,T^{n\nu}_{\text{Wak},a}(r),\qquad\qquad a=q,G,\\
\int\ud^3r\,J^{n\nu\rho}_{\text{Bel},a}(r)&=\int\ud^3r\,J^{n\nu\rho}_{\text{Ji},a}(r)=\int\ud^3r\,J^{n\nu\rho}_{\text{Wak},a}(r),\qquad\qquad a=q,G,
\end{aligned}
\end{equation}
where $J^{\mu\nu\rho}(r)=S^{\mu\nu\rho}(r)+L^{\mu\nu\rho}(r)$. 
\newline
\begin{table}[t!]
\begin{center}
\caption{\footnotesize{Expressions for the Belinfante-Rosenfeld, Ji, Wakamatsu and Chen \emph{et al.} forms  energy momentum tensors for quarks and gluons.}}\label{EMTdec}
\begin{tabular}{@{\quad}c@{\quad}|@{\quad}c@{\quad}c@{\quad}c@{\quad}|@{\quad}c@{\quad}}\whline
&Belinfante-Rosenfeld&Ji&Wakamatsu (gik) &Chen \emph{et al.} (gic)\\
\hline
$T^{\mu\nu}_q(r)$&$T^{\mu\nu}_1(r)+T^{\mu\nu}_4(r)$&$T^{\mu\nu}_1(r)$&$T^{\mu\nu}_1(r)$&$T^{\mu\nu}_1(r)+T^{\mu\nu}_3(r)$\\
$T^{\mu\nu}_G(r)$&$T^{\mu\nu}_2(r)$&$T^{\mu\nu}_2(r)$&$T^{\mu\nu}_2(r)+T^{\mu\nu}_5(r)$&$T^{\mu\nu}_2(r)-T^{\mu\nu}_3(r)+T^{\mu\nu}_5(r)$\\
\whline
\end{tabular}
\end{center}
\end{table}

\begin{table}[t!]
\begin{center}
\caption{\footnotesize{Expressions for the Belinfante-Rosenfeld, Ji, Wakamatsu and Chen \emph{et al.} forms of the generalized spin and orbital angular momentum tensors for quarks and gluons.}}\label{AMTdec}
\begin{tabular}{@{\quad}c@{\quad}|@{\quad}c@{\quad}c@{\quad}c@{\quad}|@{\quad}c@{\quad}}\whline
&Belinfante-Rosenfeld&Ji&Wakamatsu (gik) &Chen \emph{et al.} (gic)\\
\hline
$S^{\mu\nu\rho}_q(r)$&$0$&$S^{ \mu\nu\rho}_1(r)$&$S^{ \mu\nu\rho}_1(r)$&$S^{ \mu\nu\rho}_1(r)$\\
$L^{\mu\nu\rho}_q(r)$&$L^{\mu\nu\rho}_1(r)+L^{\mu\nu\rho}_4(r)$&$L^{\mu\nu\rho}_1(r)$&$L^{\mu\nu\rho}_1(r)$&$L^{\mu\nu\rho}_1(r)+L^{\mu\nu\rho}_3(r)$\\
$S^{\mu\nu\rho}_G(r)$&$0$&$0$&$S^{ \mu\nu\rho}_2(r)$&$S^{ \mu\nu\rho}_2(r)$\\
$L^{\mu\nu\rho}_G(r)$&$L^{\mu\nu\rho}_2(r)$&$L^{\mu\nu\rho}_2(r)$&$L^{\mu\nu\rho}_2(r)+L^{\mu\nu\rho}_5(r)$&$L^{\mu\nu\rho}_2(r)-L^{\mu\nu\rho}_3(r)+L^{\mu\nu\rho}_5(r)$\\
\whline
\end{tabular}
\end{center}
\end{table}

The Wakamatsu and Chen \emph{et al.} decompositions require the introduction of a pure-gauge field
\begin{equation}
A^\pure_\mu(r)\equiv\tfrac{i}{g}\,\mathcal W(r)\partial_\mu\mathcal W^{-1}(r),
\end{equation}
where $\mathcal W(r)$ (called $U_\pure(r)$ in~\cite{Lorce:2012rr}) is some phase factor transforming as $\mathcal W(r)\mapsto U(r)\mathcal W(r)$ under gauge transformations. The ``physical'' gluon field is then defined as
\begin{equation}
A^\phys_\mu(r)\equiv A_\mu(r)-A^\pure_\mu(r).
\end{equation}
In the gauge where $\mathcal W(r)=\mathds 1$, the Chen \emph{et al.} decomposition takes the same mathematical form as the Jaffe-Manohar decomposition, and can therefore be considered as a gauge-invariant extension of the latter~\cite{Ji:2012ba,Lorce:2012rr,Lorce:2013gxa,Leader:2013jra}. The phase factor $\mathcal W(r)$ is non-locally related to the field strength and is in principle not unique~\cite{Lorce:2012rr,Leader:2013jra}. The original Wakamatsu~\cite{Wakamatsu:2010qj} and Chen \emph{et al.}~\cite{Chen:2008ag} decompositions correspond to a particular choice of the phase factor which makes the physical field transverse in a given Lorentz frame. Leaving the phase factor unspecified allows us to consider at once two whole \emph{classes} of decompositions differing simply by the precise form of the non-local phase factor. In order to stress this point, we will follow from now on the terminology of Ref.~\cite{Leader:2013jra} and refer to the Wakamatsu and Chen \emph{et al.} decompositions as the gauge-invariant kinetic (gik) and canonical (gic) decompositions, respectively. For a given phase factor, the difference between the gauge-invariant kinetic and canonical decompositions lies in the separation of total linear and orbital angular momentum into quark and gluon contributions. This difference corresponds to 
\begin{equation}
\begin{aligned}
T^{\mu\nu}_3(r)&=T^{\mu\nu}_{\text{gic},q}(r)-T^{\mu\nu}_{\text{gik},q}(r)=-\left[T^{\mu\nu}_{\text{gic},G}(r)-T^{\mu\nu}_{\text{gik},G}(r)\right]\!,\\
M^{\mu\nu\rho}_3(r)&=M^{\mu\nu\rho}_{\text{gic},q}(r)-M^{\mu\nu\rho}_{\text{gik},q}(r)=-\left[M^{\mu\nu\rho}_{\text{gic},G}(r)-M^{\mu\nu\rho}_{\text{gik},G}(r)\right]\!,
\end{aligned}
\end{equation}
which are called \emph{potential} linear and angular momentum tensors~\cite{Wakamatsu:2010qj,Wakamatsu:2010cb}, respectively.

\section{Parametrization}\label{sec3}

In practice, since we want to relate the matrix elements of the gauge-invariant energy-momentum tensor to measurable parton distributions, we choose the non-local phase factor $\mathcal W(r)$ to be a Wilson line $\mathcal W_n(r,r_0)$ connecting a fixed reference point $r_0$ (usually taken at infinity) to the point of interest $r$. According to the factorization theorems~\cite{Collins:2011zzd}, these Wilson lines run essentially in a straight line along the light-front (LF) direction given by a lightlike four-vector $n$ to the intermediate point $r_n=r\pm\infty n$, and then in the transverse direction to $r_0$. In some sense, these Wilson lines can be viewed as a background gluon field generated by the hard part of the scattering. The Wilson line associated with the first part of the path
\begin{equation}\label{LFWilson}
\mathcal W_n(r,r_n)=\mathcal P\!\left[e^{-ig\int^{\pm\infty}_0n\cdot A(r+\lambda n)\,\ud \lambda}\right]
\end{equation}
is responsible for making the LF gauge $n\cdot A=0$ special, since this is the gauge where $\mathcal W_n(r,r_n)=\mathds 1$. The transverse Wilson line $\mathcal W_n(r_n,r_0)$ is associated with the residual gauge freedom and can be set to $\mathds 1$ using appropriate boundary conditions for the gauge field~\cite{Lorce:2012ce,Hatta:2011ku}. Our gauge-invariant canonical energy-momentum tensor will then be physically equivalent to the Jaffe-Manohar tensor considered in the LF gauge $n\cdot A=0$ with appropriate boundary conditions. 

We will consider in the following the generic LF gauge-invariant energy-momentum tensor of which the Belinfante-Rosenfeld, Ji, gauge-invariant kinetic and canonical energy-momentum tensors represent particular cases. The matrix elements of the generic LF gauge-invariant energy-momentum tensor depends in principle on $n$. This dependence was overlooked in~\cite{Leader:2012vp}, leading to absurd conclusions. Since any rescaled lightlike four-vector $\alpha n$ specifies the same LF Wilson line, the matrix elements of the generic LF gauge-invariant energy-momentum tensor actually depends, beside the average target momentum  $P=(p'+p)/2$ and the momentum transfer $\Delta=p'-p$, also on the following four-vector
\begin{equation}
N=\frac{M^2\,n}{P\cdot n}
\end{equation}
with $M$ the target mass, and on the parameter $\eta=\pm 1$ indicating whether the LF Wilson lines are future-pointing ($\eta=+1$) or past-pointing ($\eta=-1$). Note that, contrary to $n$, the lightlike four-vector $N$ has the same dimension and transformation properties under space-time symmetries as the momentum variables. Since $P\cdot\Delta=0$ and $M^2=P\cdot N=P^2+\Delta^2/4$, the scalar functions parametrizing the matrix elements of the generic LF gauge-invariant energy-momentum tensor are functions of the two scalar variables $\xi=-(\Delta\cdot N)/2(P\cdot N)$ and $t=\Delta^2$. Choosing the standard form for the lightlike four-vector $n=(1,0,0,-1)$ leads to the usual expression $\xi=-\Delta^+/2P^+$ with $a^\pm=a^0\pm a^3$. Because these scalar functions also depend on the parameter $\eta$, they are complex-valued just like the GTMDs~\cite{Meissner:2009ww,Lorce:2013pza}.
\newline

Using the techniques from the Appendix A of Ref.~\cite{Meissner:2009ww}, we find that the matrix elements of the generic LF gauge-invariant energy-momentum tensor for a spin-$1/2$ target can be parametrized as
\begin{equation}
\langle p',S'|T^{\mu\nu}_a(0)|p,S\rangle =\overline u(p',S')\Gamma^{\mu\nu}_a(P,\Delta,N;\eta)u(p,S),
\end{equation}
where $S$ and $S'$ are the initial and final target polarization four-vectors satisfying $p\cdot S=p'\cdot S'=0$ and $S^2=S'^2=-M^2$, and $\Gamma^{\mu\nu}_a$ stands for
\begin{equation}\label{param}
\begin{aligned}
\Gamma^{\mu\nu}_a&=Mg^{\mu\nu}A^a_1+\frac{P^\mu P^\nu}{M}\,A^a_2+\frac{\Delta^\mu\Delta^\nu}{M}\,A^a_3+\frac{P^\mu i\sigma^{\nu\Delta}}{2M}\,A^a_4+\frac{P^\nu i\sigma^{\mu\Delta}}{2M}\, A^a_5\\
&+\frac{N^\mu N^\nu}{M}\,B^a_1+\frac{P^\mu N^\nu}{M}\,B^a_2+\frac{P^\nu N^\mu}{M}\,B^a_3+\frac{N^\mu i\sigma^{\nu\Delta}}{2M}\,B^a_4+\frac{N^\nu i\sigma^{\mu\Delta}}{2M}\, B^a_5+\frac{\Delta^\mu i\sigma^{\nu N}}{2M}\,B^a_6+\frac{\Delta^\nu i\sigma^{\mu N}}{2M}\, B^a_7\\
&+\left[Mg^{\mu\nu}B^a_8+\frac{P^\mu P^\nu}{M}\,B^a_9+\frac{\Delta^\mu\Delta^\nu}{M}\,B^a_{10}+\frac{N^\mu N^\nu}{M}\,B^a_{11}+\frac{P^\mu N^\nu}{M}\,B^a_{12}+\frac{P^\nu N^\mu}{M}\,B^a_{13}\right]\frac{i\sigma^{N\Delta}}{2M^2}\\
&+\frac{P^\mu\Delta^\nu}{M}\,B^a_{14}+\frac{P^\nu \Delta^\mu}{M}\,B^a_{15}+\frac{\Delta^\mu N^\nu}{M}\,B^a_{16}+\frac{\Delta^\nu N^\mu}{M}\,B^a_{17}+\frac{M}{2}\,i\sigma^{\mu\nu}\,B^a_{18}+\frac{\Delta^\nu i\sigma^{\mu\Delta}}{2M}\, B^a_{19}\\
&+\frac{P^\mu i\sigma^{\nu N}}{2M}\,B^a_{20}+\frac{P^\nu i\sigma^{\mu N}}{2M}\, B^a_{21}+\frac{N^\mu i\sigma^{\nu N}}{2M}\,B^a_{22}+\frac{N^\nu i\sigma^{\mu N}}{2M}\, B^a_{23}\\
&+\left[\frac{P^\mu\Delta^\nu}{M}\,B^a_{24}+\frac{P^\nu \Delta^\mu}{M}\,B^a_{25}+\frac{\Delta^\mu N^\nu}{M}\,B^a_{26}+\frac{\Delta^\nu N^\mu}{M}\,B^a_{27}\right]\frac{i\sigma^{N\Delta}}{2M^2}.
\end{aligned}
\end{equation}
For convenience, we used the notation $i\sigma^{\mu b}\equiv i\sigma^{\mu\alpha}b_\alpha$. The factors of $i$ have been chosen such that the real part of the scalar functions is $\eta$-even and the imaginary part is $\eta$-odd
\begin{equation}
X^a_j(\xi,t;\eta)=X^{e,a}_j(\xi,t)+i\eta\,X^{o,a}_j(\xi,t)
\end{equation}
as a consequence of naive time-reversal symmetry. Hermiticity then implies that the real part of $B^a_j$ with $j\geq 14$ is $\xi$-odd and the imaginary part is $\xi$-even. For the other functions, the real part is $\xi$-even and the imaginary part is $\xi$-odd.

We have found that the parametrization of the matrix elements of the generic LF gauge-invariant energy-momentum tensor for a spin-$1/2$ target involves $32$ complex-valued scalar functions. This number can be obtained from a simple counting. The generic energy-momentum tensor $T^{\mu\nu}_a$ has $4\times4=16$ components. The target state polarizations $\pm S$ and $\pm S'$ bring another factor of $2\times 2=4$, but parity symmetry reduces the number of independent polarization configurations by a factor $2$, leading to a total of $32$ 
independent complex-valued amplitudes $\langle p',S'|T^{\mu\nu}_a(0)|p,S\rangle$. These 32 independent amplitudes correspond to 32 independent Dirac structures, a particular set being given by Eq.~\eqref{param}. Any other Dirac structure like \emph{e.g.} $\gamma^\mu$, $i\sigma^{\mu P}$ or $i\epsilon^{\mu\nu N\Delta}\gamma_5$, can be expressed onshell as a linear combination of these $32$ structures, see Appendix~\ref{App} of this paper.

\section{Constraints}\label{sec4}

The parametrization~\eqref{param} is very general and does not take into account several constraints like linear and angular momentum conservation. We discuss in this section the various constraints and the relation to former works on the local gauge-invariant energy-momentum tensor.  

For latter convenience, we introduce the Sudakov decomposition of a generic four-vector 
\begin{equation}
a^\mu=(a\cdot n)\bar n^\mu+(a\cdot\bar n)n^\mu+a^\mu_T.
\end{equation}
together with the transverse Kronecker and Levi-Civita symbols
\begin{equation}
\begin{aligned}
\delta^\mu_{T\nu}&=\delta^\mu_\nu-n^\mu\bar n_\nu-\bar n^\mu n_\nu,\\
\epsilon^{\mu\nu}_T&=\epsilon^{\mu\nu\alpha\beta}n_\alpha \bar n_\beta,
\end{aligned}
\end{equation}
where $\bar n$ is the lightlike four-vector satisfying $n\cdot\bar n=1$ and such that $P^\mu_T=0$.

\subsection{Local operators}

The energy-momentum tensors $T^{\mu\nu}_1(r)$ and $T^{\mu\nu}_2(r)$ are local. The corresponding matrix elements cannot therefore depend on $N$ or $\eta$. All the scalar functions must then vanish except the five real-valued functions $A^{e,a}_j(0,t)$ with $a=1,2$. These are related to the standard (local) energy-momentum form factors (FFs)~\cite{Ji:1996ek,Bakker:2004ib,Leader:2013jra} as follows
\begin{equation}\label{EMFFs}
\begin{aligned}
A_q(t)&=A^{e,1}_2(0,t),&A_G(t)&=A^{e,2}_2(0,t),\\
B_q(t)&=A^{e,1}_4(0,t)+A^{e,1}_5(0,t)-A^{e,1}_2(0,t),&\qquad B_G(t)&=A^{e,2}_4(0,t)+A^{e,2}_5(0,t)-A^{e,2}_2(0,t),\\
C_q(t)&=A^{e,1}_3(0,t),&C_G(t)&=A^{e,2}_3(0,t),\\
\bar C_q(t)&=A^{e,1}_1(0,t)+\tfrac{t}{M^2}\,A^{e,1}_3(0,t),&\bar C_G(t)&=A^{e,2}_1(0,t)+\tfrac{t}{M^2}\,A^{e,2}_3(0,t),\\
D_q(t)&=A^{e,1}_4(0,t)-A^{e,1}_5(0,t),&0&=A^{e,2}_4(0,t)-A^{e,2}_5(0,t).
\end{aligned}
\end{equation}
The first four form factors parametrize the symmetric part of the local gauge-invariant energy-momentum tensor, whereas the last one parametrizes its antisymmetric part. Since $T^{\mu\nu}_2(r)$ is symmetric, we have $A^{e,2}_4(0,t)=A^{e,2}_5(0,t)$.

\subsection{Light-front constraints}

From our choice of the phase factor~\eqref{LFWilson} it follows that $A_\phys\cdot N=0$~\cite{Hatta:2011zs,Lorce:2012ce}, leading to
\begin{equation}\label{LFcons}
T^{\mu N}_3(r)=T^{\mu N}_5(r)=0.
\end{equation}
Contracting our generic parametrization~\eqref{param} with $N_\nu$, we find the relations
\begin{equation}\label{LFconstraint}
\begin{aligned}
A^a_1(\xi,t)+B^a_3(\xi,t)-2\xi B^a_{17}(\xi,t)&=0,\\
A^a_2(\xi,t)-2\xi B^a_{14}(\xi,t)&=0,\\
-2\xi A^a_3(\xi,t)+B^a_{15}(\xi,t)&=0,\\
A^a_4(\xi,t)+B^a_9(\xi,t)-2\xi B^a_{24}(\xi,t)&=0,\\
A^a_5(\xi,t)-2\xi B^a_{19}(\xi,t)&=0,\\
B^a_4(\xi,t)+B^a_8(\xi,t)+B^a_{13}(\xi,t)-2\xi B^a_{27}(\xi,t)&=0,\\
-2\xi B^a_7(\xi,t)+B^a_{18}(\xi,t)+B^a_{21}(\xi,t)&=0,\\
-2\xi B^a_{10}(\xi,t)+B^a_{25}(\xi,t)&=0,
\end{aligned}
\end{equation}
for $a= 3,5$ which we refer to as the LF constraints.

\subsection{Four-momentum conservation}

The total energy-momentum tensor $T^{\mu\nu}(r)=T^{\mu\nu}_1(r)+T^{\mu\nu}_2(r)$ and the superpotential terms $T^{[\mu\nu]}_1(r)$ and $T^{\mu\nu}_5(r)$ are all conserved $\partial_\mu T^{\mu\nu}(r)=\partial_\mu T^{[\mu\nu]}_1(r)=\partial_\mu T^{\mu\nu}_5(r)=0$. This translates at the level of the matrix elements as
\begin{equation}
\begin{aligned}
\Delta_\mu\langle p',S'|T^{\mu\nu}(0)|p,S\rangle&=0,\\
\Delta_\mu\langle p',S'|T^{[\mu\nu]}_1(0)|p,S\rangle&=0,\\
\Delta_\mu\langle p',S'|T^{\mu\nu}_5(0)|p,S\rangle&=0,
\end{aligned}
\end{equation}
and implies the following constraints
\begin{equation}\label{consconstraint}
\begin{aligned}
\sum_{a=1}^2\left[A^{e,a}_1(0,t)+\tfrac{t}{M^2}\,A^{e,a}_3(0,t)\right]=\sum_{a=q,G}\bar C_a(t)&=0,\\
A^5_1(\xi,t)+\tfrac{t}{M^2}\,A^5_3(\xi,t)-2\xi B^5_{17}(\xi,t)&=0,\\
-2\xi B^5_1(\xi,t)+\tfrac{t}{M^2}\,B^5_{16}(\xi,t)&=0,\\
-2\xi B^5_3(\xi,t)+\tfrac{t}{M^2}\,B^5_{15}(\xi,t)&=0,\\
-2\xi B^5_4(\xi,t)-B^5_{18}(\xi,t)&=0,\\
\tfrac{t}{M^2}\,B^5_6(\xi,t)-2\xi B^5_{22}(\xi,t)&=0,\\
-B^5_7(\xi,t)+B^5_8(\xi,t)+\tfrac{t}{M^2}\,B^5_{10}(\xi,t)-2\xi B^5_{27}(\xi,t)&=0,\\
-2\xi B^5_{11}(\xi,t)-B^5_{23}(\xi,t)+\tfrac{t}{M^2}\,B^5_{26}(\xi,t)&=0,\\
-2\xi B^5_{13}(\xi,t)-B^5_{21}(\xi,t)+\tfrac{t}{M^2}\,B^5_{25}(\xi,t)&=0,
\end{aligned}
\end{equation}
which are compatible with Eq.~\eqref{LFconstraint}.

\subsection{Forward limit and momentum}

In the forward limit $\Delta\to 0$, the parametrization of the generic LF gauge-invariant energy-momentum tensor reduces to
\begin{equation}\label{forward}
\begin{aligned}
\langle P,S|T^{\mu\nu}_{a}(0)|P,S\rangle& = 2M^2g^{\mu\nu}A^{e,a}_1+2P^\mu P^\nu A^{e,a}_2+ 2N^\mu N^\nu B^{e,a}_1+2P^\mu N^\nu B^{e,a}_2+2P^\nu N^\mu B^{e,a}_3\\
&+\eta\,\epsilon^{\mu\nu SP}\,B^{o,a}_{18}-\eta\left[P^\mu \epsilon^{\nu S}_T\,B^{o,a}_{20}+P^\nu \epsilon^{\mu S}_T\,B^{o,a}_{21}+N^\mu \epsilon^{\nu S}_T\,B^{o,a}_{22}+N^\nu \epsilon^{\mu S}_T\,B^{o,a}_{23}\right]\!.
\end{aligned}
\end{equation}
Since $T^{\mu\nu}_5(r)$ is a total divergence, its matrix elements are proportional to $\Delta$ and therefore vanish in the forward limit, leading to
\begin{equation}
\begin{aligned}
A^{e,5}_1(0,0)=A^{e,5}_2(0,0)&=0,\\
B^{e,5}_1(0,0)=B^{e,5}_2(0,0)=B^{e,5}_3(0,0)&=0,\\
B^{o,5}_{18}(0,0)=B^{o,5}_{20}(0,0)=B^{o,5}_{21}(0,0)=B^{o,5}_{22}(0,0)=B^{o,5}_{23}(0,0)&=0.
\end{aligned}
\end{equation}
Moreover, since the tensors $T^{\mu\nu}_1(r)$ and $T^{\mu\nu}_2(r)$ are local, the only non-vanishing scalars $B^a_j(0,0)$ arise from the potential term $T^{\mu\nu}_3(r)$. This means in particular that naive $\mathsf T$-odd effects in the forward limit are necessarily associated with the canonical momentum and disappear when summed over all partons.
\newline

Contracting now Eq.~\eqref{forward} with $\tfrac{1}{2M^2}\,N_\mu$ gives the average four-momentum in the LF form of dynamics
\begin{equation}\label{momentum}
\langle p^\nu_a\rangle\equiv\frac{1}{2M^2}\,\langle P,S|T^{N\nu}_{a}(0)|P,S\rangle = P^\nu A^{e,a}_2+N^\nu (A^{e,a}_1+B^{e,a}_2)+\frac{\eta}{2}\,\epsilon^{\nu S}_T\,(B^{o,a}_{18}-B^{o,a}_{20}).
\end{equation}
In particular, using Eq.~\eqref{EMFFs} we recover the standard expression for the gauge-invariant kinetic four-momentum in terms of the energy-momentum FFs
\begin{equation}
\langle p^\nu_{\text{gik},a}\rangle = P^\nu A_a(0)+N^\nu \bar C_a(0),\qquad\qquad a=q,G.
\end{equation}
Interestingly, the last term in Eq.~\eqref{momentum} is naive $\mathsf T$-odd and can be interpreted as the spin-dependent contribution to the momentum arising from initial and/or final-state interactions. Because of the structure $\epsilon^{\nu S}_T$, this naive $\mathsf T$-odd contribution is transverse and requires a transverse target polarization. As we will see in section~\ref{sec5b}, this is related to the Sivers effect~\cite{Sivers:1989cc}. The combination of scalars $A^{e,a}_1(0,0)+B^{e,a}_2(0,0)$ contributes only to the energy and is therefore related to the interaction term in the Hamiltonian. In the forward limit, the LF constraints~\eqref{LFconstraint} imply that 
\begin{equation}
A^{e,3}_2(0,0)=0,
\end{equation}
unless $B^{e,3}_{14}(\xi,t)$ behaves as $1/\xi$ near $\xi=0$. This suggests that the scalars $A^{e,1}_2(0,0)$ and $A^{e,2}_2(0,0)$, and hence $A_q(0)$ and $A_G(0)$, can be interpreted as the interaction-independent contributions of, respectively, quarks and gluons to the four-momentum. This is further supported by the observation that $A^{e,a}_2(0,0)$ is the only contribution to the longitudinal momentum $\langle p^n_a\rangle$, which is purely kinematical in LF quantization. In other words, there is no difference between the longitudinal component of the average kinetic and canonical momenta.

Finally, since the total four-momentum is $\langle p^\nu\rangle= P^\nu$, we obtain from Eq.~\eqref{momentum} the momentum constraints
\begin{equation}\label{momcons}
\begin{aligned}
\sum_{a=1,2}A^{e,1}_1(0,0)=\sum_{a=q,G}\bar C_a(0)&=0,\\
\sum_{a=1,2}A^{e,1}_2(0,0)=\sum_{a=q,G} A_a(0)&=1,
\end{aligned}
\end{equation}
which are consistent with Eq.~\eqref{consconstraint}. In particular, the vanishing of the average total transverse momentum, known as the Burkardt sum rule~\cite{Burkardt:2003yg,Burkardt:2004ur}, is trivially taken into account in our parametrization because the potential term $T^{\mu\nu}_3(r)$, and hence $B^{o,3}_{18}(0,0)-B^{o,3}_{20}(0,0)$, drops out of the sum over all partons.

\subsection{Angular momentum}\label{sec4e}

Since we have a complete parametrization of the matrix elements of the generic LF gauge-invariant energy-momentum tensor, we can easily compute the matrix elements of the corresponding OAM tensor $L^{\mu\nu\rho}_{a}(r)$ given by Eq.~\eqref{AMtensors}. Because of the explicit factors of position $r$, the matrix elements of the generic LF gauge-invariant OAM tensor need to be handled with care~\cite{Bakker:2004ib,Leader:2013jra}. Focusing on the longitudinal component of OAM, we find
\begin{align}
\langle L^a_L\rangle\equiv\frac{1}{2M^2}\,\langle P,S|\tfrac{1}{2}\,\epsilon_{T\alpha\beta}L^{N\alpha\beta}_{a}(0)|P,S\rangle&=\frac{\epsilon_{T\alpha\beta}}{2M^2}\left[i\,\frac{\partial}{\partial\Delta_\alpha}\langle p',S'|T^{N \beta}_{a}(0)|p,S\rangle\right]_{\Delta=0}\nonumber\\
&= \frac{S\cdot N}{M^2}\,A^{e,a}_4(0,0).\label{OAM}
\end{align}
For a longitudinally polarized target, we have $S\cdot N=M^2$ and so $A^{e,a}_4(0,0)$ can be interpreted as the average fraction of target longitudinal angular momentum carried by the OAM associated with the energy-momentum tensor $T^{\mu\nu}_a(r)$ in the LF form of dynamics. We confirm in particular that the integrated OAM does not receive any naive $\mathsf T$-odd contribution~\cite{Hatta:2011ku,Hatta:2012cs,Lorce:2012ce,Ji:2012ba}. Using Eq.~\eqref{EMFFs}, we also recover the standard expressions for the Belinfante and Ji forms of longitudinal OAM in terms of the energy-momentum FFs~\cite{Ji:1996ek,Shore:1999be,Bakker:2004ib,Leader:2013jra}
\begin{equation}\label{AMtot}
\begin{aligned}
\langle J^{a}_{\text{Bel},L}\rangle=\langle L^{a}_{\text{Bel},L}\rangle&=\tfrac{1}{2}\left[A_a(0)+B_a(0)\right]\tfrac{S\cdot N}{M^2},\qquad\qquad\qquad a=q,G,\\
\langle L^{q}_{\text{Ji},L}\rangle&=\tfrac{1}{2}\left[A_q(0)+B_q(0)+D_q(0)\right]\tfrac{S\cdot N}{M^2}.
\end{aligned}
\end{equation}


Remarkably, thanks to Eq.~\eqref{antisymm} we can also express the quark and gluon spin contributions in terms of the scalar functions parametrizing the generic LF gauge-invariant energy-momentum tensor. From the integral relations in Eq.~\eqref{antisymm}, we find that the quark and gluon longitudinal spin contributions are given by
\begin{equation}\label{spin}
\begin{aligned}
\langle S^q_L\rangle\equiv\frac{1}{2M^2}\,\langle P,S|\tfrac{1}{2}\,\epsilon_{T\alpha\beta}S^{N\alpha\beta}_1(0)|P,S\rangle&=-\frac{1}{2}\left[A^{e,1}_4(0,0)-A^{e,1}_5(0,0)\right]\frac{S\cdot N}{M^2},\\
\langle S^G_L\rangle\equiv\frac{1}{2M^2}\,\langle P,S|\tfrac{1}{2}\,\epsilon_{T\alpha\beta}S^{N\alpha\beta}_2(0)|P,S\rangle&=-\frac{S\cdot N}{M^2}\,A^{e,5}_4(0,0).
\end{aligned}
\end{equation}
where we have used Eq.~\eqref{QCDEOM} to express $L^{\mu\nu\rho}_4(r)$ in terms of $T^{\mu\nu}_1(r)$. The scalars $-\tfrac{1}{2}[A^{e,1}_4(0,0)-A^{e,1}_5(0,0)]=-\tfrac{1}{2}D_q(0)$ and  $-A^{e,5}_4(0,0)$ can therefore be interpreted as the average fraction of target longitudinal angular momentum carried by the spin of quarks and gluons, respectively. It is easy to check that the following relations
\begin{equation}
\langle S^a_L\rangle+\langle L^{a}_{\text{gik},L}\rangle=\langle J^{a}_{\text{Bel},L}\rangle,\qquad\qquad a=q,G
\end{equation}
are satisfied and that the longitudinal component of the potential OAM is given by
\begin{equation}
\langle L^{q}_{\text{gic},L}\rangle-\langle L^{q}_{\text{gik},L}\rangle=-\left[\langle L^{G}_{\text{gic},L}\rangle-\langle L^{G}_{\text{gik},L}\rangle\right]=\frac{S\cdot N}{M^2}\,A^{e,3}_4(0,0).
\end{equation}
Note that the differential relations in Eq.~\eqref{antisymm}, which translate at the level of matrix elements as
\begin{equation}\label{spinidentity}
\begin{aligned}
\langle p',S'|T^{[\nu\rho]}_1(0)|p,S\rangle&=-i\Delta_\mu\langle p',S'|S^{\mu\nu\rho}_1(0)|p,S\rangle,\\
\langle p',S'|T^{[\nu\rho]}_5(0)|p,S\rangle
&=-i\Delta_\mu\langle p',S'|S^{\mu\nu\rho}_2(0)|p,S\rangle,
\end{aligned}
\end{equation}
do not provide additional constraints. Indeed, at $\mathcal O(\Delta^0)$, they just reduce to the antisymmetric part of the forward limit~\eqref{forward}. At higher orders in $\Delta$, the identification of coefficients between the LHS and the RHS of Eq.~\eqref{spinidentity}  are spoiled by the condition $P\cdot\Delta=0$ which follows from the onshell relation for the target $(P\pm\tfrac{\Delta}{2})^2=M^2$.

Finally, since the total angular momentum is $1/2$, we obtain from Eqs.~\eqref{OAM} and~\eqref{spin} the angular momentum constraints
\begin{equation}\label{cons3}
\sum_{a=1,2}[A^{e,a}_4(0,0)+A^{e,a}_5(0,0)]=\sum_{a=q,G}\left[A_a(0)+B_a(0)\right]=1.
\end{equation}
Combined with the momentum constraints~\eqref{momcons}, this leads to
\begin{equation}\label{cons4}
\sum_{a=1,2}[A^{e,a}_4(0,0)+A^{e,a}_5(0,0)-A^{e,a}_2(0,0)]=\sum_{a=q,G}B_a(0)=0.
\end{equation}
known as the anomalous gravitomagnetic moment sum rule~\cite{Teryaev:1999su,Brodsky:2000ii}.

\section{Link with measurable parton distributions}\label{sec5}

Now we are going to see how the scalar functions parametrizing the matrix elements of the generic LF gauge-invariant energy-momentum tensor are related to GPDs accessed in exclusive scatterings~\cite{Diehl:2003ny} and TMDs accessed in semi-inclusive scatterings~\cite{Collins:2011zzd}. For convenience, we shall focus in the following on the quark sector. The gluon sector proceeds analogously.

\subsection{Generalized Parton Distributions}

The quark vector GPD correlator is defined as
\begin{equation}\label{GPDcorr}
F^{[\gamma^\mu]}_{S'S}(P,x,\Delta,N)=(P\cdot n)\int\frac{\ud\lambda}{2\pi}\,e^{ix\lambda (P\cdot n)}\,\langle p',S'|\barpsi(-\tfrac{\lambda}{2}n)\gamma^\mu\mathcal W_n(-\tfrac{\lambda}{2}n,\tfrac{\lambda}{2}n)\psi(\tfrac{\lambda}{2}n)|p,S\rangle.
\end{equation}
Remarkably, its second Mellin moment is related to the matrix elements of the quark LF gauge-invariant energy-momentum tensor~\cite{Diehl:2003ny,Lorce:2012ce,Leader:2013jra}
\begin{equation}\label{GPDrel}
\int\ud x\,x\,F^{[\gamma^\mu]}_{S'S}(P,x,\Delta,N)=\tfrac{1}{M^2}\,\langle p',S'|T^{\mu N}_{q}(0)|p,S\rangle.
\end{equation}
Note that we do need to specify whether this corresponds to the kinetic or canonical version of the LF gauge-invariant energy-momentum tensor simply because $T^{\mu N}_{\text{gik},q}(r)=T^{\mu N}_{\text{gic},q}(r)$ owing to Eq.~\eqref{LFcons}. 

Up to twist 4, the quark vector GPD correlator~\eqref{GPDcorr} is parametrized as~\cite{Meissner:2009ww}
\begin{equation}
\begin{aligned}
F^{[\uslash n]}_{S'S}(P,x,\Delta,N)&=\overline u(p',S')\!\left[\uslash n H^q+\frac{i\sigma^{n\Delta}}{2M}\,E^q\right]\!u(p,S),\\
F^{[\gamma^\mu_T]}_{S'S}(P,x,\Delta,N)&=\overline u(p',S')\!\left[\frac{Mi\sigma^{n\alpha}_T}{P\cdot n}\,H^q_{2T}+\frac{\uslash n\Delta^\alpha_T}{2(P\cdot n)}\,E^q_{2T}+\frac{\Delta^\alpha_T}{M}\,\tilde H^q_{2T}-\gamma^\alpha_T\left(\tilde E^q_{2T}-\xi E^q_{2T}\right)\right]\!u(p,S),\\
F^{[\uslash \bar n]}_{S'S}(P,x,\Delta,N)&=\frac{M^2}{(P\cdot n)^2}\,\overline u(p',S')\!\left[\uslash n H^q_3+\frac{i\sigma^{n\Delta}}{2M}\,E^q_3\right]\!u(p,S).
\end{aligned}
\end{equation}
From Eq.~\eqref{GPDrel}, we then find the relations between the second Mellin moment of vector GPDs and the energy-momentum FFs in the quark sector
\begin{equation}\label{poly}
\begin{split}
\int\ud x\,x\,H^q(x,\xi,t)&=A_q(t)+4\xi^2\,C_q(t),\\
\int\ud x\,x\,E^q(x,\xi,t)&=B_q(t)-4\xi^2\,C_q(t),\\
\int\ud x\,x\,H^q_{2T}(x,\xi,t)&=0,\\
\int\ud x\,x\,E^q_{2T}(x,\xi,t)&=0,\\
\int\ud x\,x\,\tilde H^q_{2T}(x,\xi,t)&=-2\xi\, C_q(t),\\
\int\ud x\,x\,\tilde E^q_{2T}(x,\xi,t)&=-\tfrac{1}{2}\!\left[A_q(t)+B_q(t)-D_q(t)\right]\!,\\
\int\ud x\,x\,H^q_3(x,\xi,t)&=\tfrac{1}{2}A_q(t)+\bar C_q(t)-2\xi^2\tfrac{P^2}{M^2}\,C_q(t)+\tfrac{t}{8M^2}\left[B_q(t)-8C_q(t)-D_q(t)\right]\!,\\
\int\ud x\,x\left[H^q_3(x,\xi,t)+E^q_3(x,\xi,t)\right]&=\tfrac{P^2}{2M^2}\,D_q(t).
\end{split}
\end{equation}
The relations involving twist-3 GPDs are consistent with those found in Ref.~\cite{Kiptily:2002nx} where the parametrization is related to the one we used as follows
\begin{equation}
\begin{split}
H^q_{2T}(x,\xi,t)&=-2\xi\,G^q_4(x,\xi,t),\\
E^q_{2T}(x,\xi,t)&=2\left[G^q_3(x,\xi,t)-\xi G^q_4(x,\xi,t)\right]\!,\\
\tilde H^q_{2T}(x,\xi,t)&=\tfrac{1}{2}G^q_1(x,\xi,t),\\
H^q(x,\xi,t)+E^q(x,\xi,t)+\tilde E^q_{2T}(x,\xi,t)&=-G^q_2(x,\xi,t)+2\!\left[\xi G^q_3(x,\xi,t)-G^q_4(x,\xi,t)\right]\!.
\end{split}
\end{equation}
The explicit relations involving twist-4 GPDs are new but somewhat academical as these functions are much harder to access experimentally\footnote{We note in passing a typo in  Ref.~\cite{Leader:2013jra} where a factor $\tfrac{1}{2}$ is missing in front of the $D_q(t)$ energy-momentum FF in the RHS of Eq.~(432).}. 

Thanks to Eq.~\eqref{spinidentity}, the antisymmetric part of the LF gauge-invariant kinetic energy-momentum tensor can be related to the local axial-vector correlator 
\begin{equation}
-\tfrac{i}{2}\,\epsilon^{\mu\nu\Delta\alpha}\int\ud x\,F^{[\gamma_\alpha\gamma_5]}_{S'S}(P,x,\Delta,N)=\langle p',S'|T^{[\mu\nu]}_{\text{gik},q}(0)|p,S\rangle.
\end{equation}
From the parametrization~\cite{Diehl:2003ny}
\begin{equation}
\int\ud x\,F^{[\gamma^\mu\gamma_5]}_{S'S}(P,x,\Delta,N)=\overline u(p',S')\!\left[\gamma^\mu\gamma_5\, G^q_A(t)+\frac{\Delta^\mu\gamma_5}{2M}\, G^q_P(t)\right]\!u(p,S),
\end{equation}
where $G^q_A(t)=\int\ud x \,\tilde H^q(x,\xi,t)$ is the axial-vector FF and $G^q_P(t)=\int\ud x \,\tilde E^q(x,\xi,t)$ is the induced pseudoscalar FF, it is easy to show using onshell identities that~\cite{Shore:1999be,Bakker:2004ib,Leader:2013jra}
\begin{equation}
G^q_A(t)=-D_q(t).
\end{equation}
This is naturally consistent with the observation in section~\ref{sec4e} that the scalar $-\tfrac{1}{2}D_q(0)$ can be regarded as the quark spin contribution to the total angular momentum in a longitudinally polarized target.

\subsection{Transverse-Momentum dependent Distributions}\label{sec5b}

The quark vector TMD correlator is defined as\footnote{For simplicity, we considered the naive definition of the TMD correlator where the soft factor is not included~\cite{Collins:2011zzd}. This allows one to treat in a simple way the $k_T$-integrations. A more careful treatment based on the proper definition of TMDs proceeds analogously.}
\begin{equation}\label{TMDcorr}
\Phi^{[\gamma^\mu]}_{S'S}(P,x,k_T,N;\eta)=(P\cdot n)\int\ud (k\cdot\bar n)\int\frac{\ud^4z}{(2\pi)^4}\,e^{ik\cdot z}\,\langle p,S'|\barpsi(-\tfrac{z}{2})\gamma^\mu\mathcal W_n(-\tfrac{z}{2},\tfrac{z}{2})\psi(\tfrac{z}{2})|p,S\rangle.
\end{equation}
Similarly to the GPD case, the second Mellin moment of this correlator is related to the forward matrix elements of the quark LF gauge-invariant energy-momentum tensor
\begin{equation}\label{TMDrel1}
\int\ud x\,\ud^2k_T\,x\,\Phi^{[\gamma^\mu]}_{S'S}(P,x,k_T,N;\eta)=\tfrac{1}{M^2}\,\langle p,S'|T^{\mu N}_{q}(0)|p,S\rangle.
\end{equation}
Remarkably, treating with due care the LF Wilson line~\cite{Boer:2003cm,Bomhof:2007xt,Buffing:2011mj,Buffing:2012sz,Hatta:2011ku,Lorce:2012ce,Leader:2013jra},  one can similarly show that the second transverse moment of $\Phi^{[\gamma^\mu]}_{S'S}$ is related to the forward matrix elements of the quark LF gauge-invariant \emph{canonical} energy-momentum tensor
\begin{equation}\label{TMDrel2}
\int\ud x\,\ud^2k_T\,k^\alpha_T\,\Phi^{[\gamma^\mu]}_{S'S}(P,x,k_T,N;\eta)=\delta^\alpha_{T\nu}\,\langle p,S'|T^{\mu\nu}_{\text{gic},q}(0)|p,S\rangle.
\end{equation}
That the relation holds for the canonical version of the LF gauge-invariant energy-momentum tensor is determined by the particular shape~\eqref{LFWilson} of the Wilson line. Working instead with a straight Wilson line connecting directly the points $\pm\tfrac{z}{2}$, the relation~\eqref{TMDrel2} would hold for the kinetic version of the LF gauge-invariant energy-momentum tensor~\cite{Ji:2012sj,Lorce:2012ce,Burkardt:2012sd}. We stress once again that our choice for the Wilson line~\eqref{LFWilson} was simply motivated by the fact that factorization theorems require Wilson lines that run essentially along the LF direction $n$~\cite{Collins:2011zzd}.

Up to twist 4, the quark vector TMD correlator~\eqref{TMDcorr} is parametrized as~\cite{Boer:1997nt,Bacchetta:2006tn,Meissner:2009ww}
\begin{equation}
\begin{aligned}
\Phi^{[\uslash n]}_{SS}(P,x,k_T,N;\eta)&=2(P\cdot n)\!\left[f^q_1-\eta\,\tfrac{\epsilon^{kS}_T}{M^2}\,f^{\perp q}_{1T}\right]\!,\\
\Phi^{[\gamma^\mu_T]}_{SS}(P,x,k_T,N;\eta)&=2M\left[\tfrac{k^\mu_T}{M}\,f^{\perp q}-\eta\,\tfrac{\epsilon^{\mu S}_T}{M}\,f^q_T-\eta\,\tfrac{(k^\mu_Tk_{T\nu}-\tfrac{1}{2}\,\delta^\mu_{T\nu}k^2_T)\,\epsilon^{\nu S}_T}{M^3}\,f^{\perp q}_T-\eta\,\tfrac{(S\cdot n)\,\epsilon^{\mu k}_T}{(P\cdot n)M}\,f^{\perp q}_L\right]\!,\\
\Phi^{[\uslash \bar n]}_{SS}(P,x,k_T,N;\eta)&=\frac{2M^2}{P\cdot n}\!\left[f^q_3-\eta\,\tfrac{\epsilon^{kS}_T}{M^2}\,f^{\perp q}_{3T}\right]\!,
\end{aligned}
\end{equation}
where we have extracted explicitly the $\eta$-dependence. From Eq.~\eqref{TMDrel1} and the forward limit $\Delta\to 0$ of the LF constraints~\eqref{LFconstraint}, we find the relations between the second Mellin moment of vector TMDs and the energy-momentum FFs in the quark sector
\begin{equation}\label{relTMD1}
\begin{split}
\int\ud x\,\ud^2k_T\,x\,f^q_1(x,k^2_T)&=A_q(0),\\
\int\ud x\,\ud^2k_T\,x\,f^q_T(x,k^2_T)&=0,\\
\int\ud x\,\ud^2k_T\,x\,f^q_3(x,k^2_T)&=\tfrac{1}{2}A_q(0)+\bar C_q(0).
\end{split}
\end{equation}
Together with Eq.~\eqref{poly}, these relations are consistent with the fact that the GPD and TMD correlators have the same collinear forward limit
\begin{equation}
\int\ud^2k_T\,\Phi^{[\gamma^\mu]}_{SS}(P,x,k_T,N;\eta)=F^{[\gamma^\mu]}_{SS}(P,x,\Delta=0,N)
\end{equation}
which implies
\begin{equation}
\begin{aligned}
\int\ud^2k_T\, f^q_1(x,k^2_T)&=H^q(x,0,0),\\
\int\ud^2k_T\, f^q_T(x,k^2_T)&=0=H^q_{2T}(x,0,0),\\
\int\ud^2k_T\, f^q_3(x,k^2_T)&=H^q_3(x,0,0).
\end{aligned}
\end{equation}

More interesting are the relations involving the second transverse moment of TMDs. From Eq.~\eqref{TMDrel2} and the forward limit $\Delta\to 0$ of the LF constraints~\eqref{LFconstraint}, we find
\begin{equation}\label{relTMD2}
\begin{split}
\int\ud x\,\ud^2k_T\,\tfrac{k^2_T}{2M^2}\,f^{\perp q}_{1T}(x,k^2_T)&=-\tfrac{1}{2}\!\left[B^{o,3}_{18}(0,0)-B^{o,3}_{20}(0,0)\right]\!,\\
\int\ud x\,\ud^2k_T\,\tfrac{k^2_T}{2M^2}\,f^{\perp q}(x,k^2_T)&=\bar C_q(0)+A^{e,3}_1(0,0),\\
\int\ud x\,\ud^2k_T\,\tfrac{k^2_T}{2M^2}\,f^{\perp q}_L(x,k^2_T)&=\tfrac{1}{2}B^{o,3}_{18}(0,0),\\
\int\ud x\,\ud^2k_T\,\tfrac{k^2_T}{2M^2}\,f^{\perp q}_{3T}(x,k^2_T)&=\tfrac{1}{4}\!\left[B^{o,3}_{18}(0,0)+B^{o,3}_{20}(0,0)+2B^{o,3}_{22}(0,0)\right]\!.
\end{split}
\end{equation}
Anticipating that similar results hold in the gluon sector, we sum over all partons and obtain the following sum rules
\begin{equation}\label{newSR}
\begin{split}
\sum_{a=q,G}\int\ud x\,\ud^2k_T\,\tfrac{k^2_T}{2M^2}\,f^{\perp a}_{1T}(x,k^2_T)&=0,\\
\sum_{a=q,G}\int\ud x\,\ud^2k_T\,\tfrac{k^2_T}{2M^2}\,f^{\perp a}(x,k^2_T)&=0,\\
\sum_{a=q,G}\int\ud x\,\ud^2k_T\,\tfrac{k^2_T}{2M^2}\,f^{\perp a}_L(x,k^2_T)&=0,\\
\sum_{a=q,G}\int\ud x\,\ud^2k_T\,\tfrac{k^2_T}{2M^2}\,f^{\perp a}_{3T}(x,k^2_T)&=0.
\end{split}
\end{equation}
The first sum rule is known as the Burkardt sum rule~\cite{Burkardt:2003yg,Burkardt:2004ur} and simply expresses the fact that the total momentum transverse to the target momentum has to vanish. The other three sum rules are to the best of our knowledge new. They express the fact that the total \emph{flow} of transverse momentum has also to vanish. They involve higher-twist TMDs and are therefore much harder to test experimentally. Nevertheless, it would be very interesting to test them using phenomenological models, Lattice QCD and perturbative QCD.

As a final remark, we would like to stress that the above results show explicitly that TMDs cannot provide any information about the scalar $A^{e,a}_4(0,0)$, which means no quantitative model-independent information about the parton OAM, as anticipated \emph{e.g.} in~\cite{Lorce:2011kn}



\section{Conclusions}\label{sec6}

There has been a prejudice against the canonical form of the quark and gluon energy-momentum tensor, and consequently of the corresponding linear and orbital angular momenta, due to the fact that it cannot be written locally in a gauge-invariant way. A gauge-invariant expression can however be obtained by relaxing the locality requirement in a way that does not harm causality. This indicates that the canonical energy-momentum tensor \emph{can} be considered as a physical object and measured experimentally. In particular, it can be accessed \emph{via} particular moments of two- and three-parton correlators which are extracted from numerous physical processes.

In this study, we provided for the first time a complete parametrization for the matrix elements of the generic asymmetric, non-local and gauge-invariant canonical energy-momentum tensor. We found that a generic canonical energy-momentum tensor for a spin-$1/2$ target consists in 32 independent complex amplitudes. We discussed in detail the various constraints on these amplitudes imposed by non-locality, linear and angular momentum conservation. This generalizes therefore former works on the symmetric, local and gauge-invariant kinetic energy-momentum tensor also known as the Belinfante-Rosenfeld energy-momentum tensor.

We also showed that some of the amplitudes can be expressed in terms of particular moments of two-parton generalized and transverse-momentum dependent distributions, and are therefore clearly measurable. In particular, we proved explicitly that two-parton transverse-momentum dependent distributions cannot provide any quantitative model-independent information about the parton orbital angular momentum. On the way, we recovered the Burkardt sum rule, expressing basically conservation of transverse momentum, and derived three new sum rules invvolving higher-twist distributions. We obtained these results by choosing the non-local phase factors defined by a lightlike four-vector $n$, in order to make contact with parton physics and factorization theorems. 

We believe the present paper will help clarify the differences between canonical and kinetic  energy-momentum tensors, and their links with parton distributions. We also expect getting more insights into these matters in a near future coming from explicit results obtained within covariant models, Lattice QCD and perturbative QCD.

\section*{Acknowledgements}

I am thankful to S. Brodsky, E.~Leader, B.~Pasquini and P. Schweitzer for useful discussions related to this study. This work was supported by the Belgian Fund F.R.S.-FNRS \emph{via} the contract of Charg\'e de Recherches.

\appendix

\section{Parametrization}\label{App}

From the discrete space-time symmetries, we find that the Dirac structure $\Gamma^{\mu\nu}_a$ associated with the matrix elements of the generic LF gauge-invariant energy-momentum tensor has to satisfy the following constraints
\begin{equation}
\begin{aligned}
\Gamma^{\mu\nu}_a(P,\Delta,N;\eta)&=\gamma^0\Gamma^{\dag\mu\nu}_a(P,-\Delta,N;\eta)\gamma^0&&\qquad\text{Hermiticity}\\
&=\gamma^0\Gamma^{\bar\mu\bar\nu}_a(\bar P,\bar \Delta,\bar N;\eta)\gamma^0&&\qquad\text{Parity}\\
&=(-i\gamma_5C)\Gamma^{*\bar\mu\bar\nu}_a(\bar P,\bar \Delta,\bar N;-\eta)(-i\gamma_5C)&&\qquad\text{Time-reversal}\\
\end{aligned}
\end{equation}
where $C$ is the charge conjugation matrix and $\bar b=b^{\bar \mu}=(b^0,-\vec b)$. Using the Gordon identities
\begin{equation}
\begin{aligned}
\overline u(p',S')\gamma^\mu u(p,S)&=\overline u (p',S')\left[\frac{P^\mu}{M}+\frac{i\sigma^{\mu\Delta}}{2M}\right]u(p,S),\\
0&=\overline u (p',S')\left[\frac{\Delta^\mu}{2M}+\frac{i\sigma^{\mu P}}{M}\right]u(p,S),
\end{aligned}
\end{equation}
we can discard the $\gamma^\mu$ and $i\sigma^{\mu P}$ structures from the parametrization. Similarly,  we can discard the structure $\epsilon^{\mu\nu\alpha P}\gamma_5$ thanks to the following onshell identity
\begin{equation}\label{id5}
\overline u(p',S')\left[2\epsilon^{\mu\nu\alpha P}\gamma_5+\Delta^\mu\sigma^{\nu\alpha}+\Delta^\nu\sigma^{\alpha\mu}+\Delta^\alpha\sigma^{\mu\nu}\right]u(p,S)=0,
\end{equation}
Contracting the $\epsilon$ identity
\begin{equation}
g^{\alpha\beta}\epsilon^{\mu\nu\rho\sigma}+g^{\alpha\mu}\epsilon^{\nu\rho\sigma\beta}+g^{\alpha\nu}\epsilon^{\rho\sigma\beta\mu}+g^{\alpha\rho}\epsilon^{\sigma\beta\mu\nu}+g^{\alpha\sigma}\epsilon^{\beta\mu\nu\rho}=0
\end{equation}
with $P_\beta\Delta_\rho N_\sigma$ gives
\begin{equation}
P^\alpha\epsilon^{\mu\nu \Delta N}+g^{\alpha[\mu}\epsilon^{\nu]\Delta NP}+\Delta^\alpha\epsilon^{\mu\nu NP}-N^\alpha\epsilon^{\mu\nu\Delta P}=0.
\end{equation} 
Contracting further with $N_\alpha$, $P_\alpha$ and $\Delta_\alpha$ leads to
\begin{equation}
\begin{aligned}
(P\cdot N)\,\epsilon^{\mu\nu\Delta N}+N^{[\mu}\epsilon^{\nu]\Delta NP}+(\Delta\cdot N)\,\epsilon^{\mu\nu NP}&=0,\\
P^2\,\epsilon^{\mu\nu\Delta N}+P^{[\mu}\epsilon^{\nu]\Delta NP}-(P\cdot N)\,\epsilon^{\mu\nu \Delta P}&=0,\\
\Delta^{[\mu}\epsilon^{\nu]\Delta NP}+\Delta^2\,\epsilon^{\mu\nu NP}-(\Delta\cdot N)\,\epsilon^{\mu\nu \Delta P}&=0.
\end{aligned}
\end{equation}
Multiplying now by $\overline u(p',S')\gamma_5 u(p,S)$ and using Eq.~\eqref{id5}, the first two identities allow us to discard the structures $\epsilon^{\mu\nu\Delta N}\gamma_5$ and $\Delta^{[\mu}\sigma^{\nu]\Delta}$ while the last identity is trivially satisfied. We are then left with the 32 independent structures given in Eq.~\eqref{param}.

\end{document}